\documentclass[12pt]{iopart}
\usepackage{iopams}
\usepackage{graphicx}
\usepackage{caption}
\usepackage{subcaption}
\usepackage{hyperref}
\usepackage{hypernat}
\usepackage{color}
\usepackage{cite}

\begin{document}
	
\title[Antifragility in the synchronization of oscillators on networks]{Antifragility and response to damage in the synchronization of oscillators on networks}
\author{M. A. Polo-González${}^{1}$, A. P. Riascos${}^{1}$\footnote{E-mail Corresponding Author: alperezri@unal.edu.co} , L. K. Eraso-Hernandez${}^2$}
	
\address{${}^{1}$Departamento de F\'isica, Universidad Nacional de Colombia, Bogotá, Colombia\\
	${}^2$Instituto de Física, Universidad Nacional Autónoma de México, Ciudad Universitaria, 04510, Mexico City, Mexico}

\date{\today}

\begin{abstract}
In this paper, we introduce a mathematical framework to assess the impact of damage, defined as the reduction of weight in a specific link, on identical oscillator systems governed by the Kuramoto model and coupled through weighted networks. We analyze how weight modifications in a single link affect the system when its global function is to achieve the synchronization of coupled oscillators starting from random initial phases. We introduce different measures that allow the identification of cases where damage enhances synchronization (antifragile response), deteriorates it (fragile response), or has no significant impact.
Using numerical solutions of the Kuramoto model, we investigate the effects of damage on network links where antifragility emerges. Our analysis includes lollipop graphs of varying sizes and a comprehensive evaluation and all the edges of 109 non-isomorphic graphs with six nodes. The approach is general and can be applied to study antifragility in other oscillator systems with different coupling mechanisms, offering a pathway for the quantitative exploration of antifragility in diverse complex systems.

\end{abstract}

\begin{titlepage}
\maketitle
\end{titlepage}
\section{Introduction}
Complex systems consist of many interconnected components and are highly sensitive to perturbations \cite{BaryamBook_1997,SayamaBook}. Such systems are often vulnerable to various types of failures, which can lead to a reduction in their global functionality in response to damage \cite{VespiBook,Carlson_PRL2000,Carlson_PNAS_2002,Dobson2007,SunPNAS2020,VuralPRE2014,Cohen2016}. In this context, studying dynamical processes in systems with damage accumulation is crucial for identifying vulnerabilities and achieving a deeper understanding of the relationship between a system's dynamics and its underlying structure \cite{VespiBook,barabasi2016book}. Recent research has focused on identifying optimal structures for transport processes \cite{Katifori_PRL2010}, strategies for network repair following attacks \cite{Farr_PRL_2014,Hu2016,Lin2020}, and understanding how damage influences diffusive dynamics \cite{Aging_PhysRevE2019,Eraso_Hernandez_2021,Eraso-metro}. Additional studies have addressed the synchronization of oscillators on networks \cite{Eraso-Hernandez_2023}, as well as the response of interacting neuronal units to targeted attacks or failures \cite{FaciLazaro2022,TellerENEURO2020,mi13122259}.
\\[2mm]
Furthermore, some systems exhibit robustness \cite{Carlson_PRL2000}, meaning they remain largely unaffected by damage. At the other extreme, fragile systems can break or collapse when subjected to damage. In this context, a system is considered antifragile if it benefits from damage, where damage acts as a constructive element that enhances the system’s capacity to perform its global function. The concept of antifragility was introduced by Taleb in his book \cite{Taleb_book} and has since been explored in various domains, including financial systems, stocks, and cryptocurrencies \cite{Stocks_Alatorre2023}, renewable energy system design \cite{Coppitters2023}, ecosystems \cite{Equihua_PeerJ_2020}, and many others \cite{axenie2023antifragility}. A similar phenomenon has been observed in simulations of urban road networks, where blocking certain streets can unexpectedly improve overall traffic conditions by reducing congestion \cite{Hyejin_PRL2008}. However, despite its relevance in understanding the effects of damage in complex systems and its potential applications across various fields, antifragility remains poorly characterized, and the fundamental conditions under which it emerges are not yet fully understood \cite{axenie2023antifragility}. Recent efforts to address these gaps include modeling antifragility in Boolean networks \cite{Pineda2019,Lopez_Entropy_2023}, the study of strongly coupled Coulomb fluids \cite{Ghodrat_2015} and emergence of antifragility in the movement of random walkers on networks with modular structures or communities \cite{Eraso_PRE_2024}.
\\[2mm]
In this paper, we investigate the emergence of antifragility in the synchronization of oscillators coupled through weighted networks. Synchronization is a collective emergent process in which a set of coupled agents, under certain conditions, self-organize and evolve to follow a common dynamical pattern \cite{BOCCALETTI2008,PikoBook,strogatzbook,J_Kurths_PhysRep2023}. This phenomenon is one of the most compelling topics in the study of complex systems, with applications ranging from the synchronization of flashing fireflies \cite{PikoBook,doi:10.1126/sciadv.abg9259} and crowd clapping at massive events \cite{neda}, to synchronization in arrays of Josephson junctions in condensed matter physics \cite{josephson}. Given its ubiquity, synchronization plays a fundamental role in the dynamics and functionality of diverse natural and technological systems \cite{PikoBook,strogatzbook,Balanov2009}. In particular, the Kuramoto model is a paradigmatic framework for studying collective systems composed of agents interacting through pairwise nonlinear couplings. Originally introduced to describe chemical instabilities \cite{Yoshiki_K}, it has become a standard model for investigating the transition to synchrony in agent-based systems. Its versatility has led to applications across a wide range of complex systems represented by networks, including neuroscience, ecology, and the humanities, among many others \cite{BOCCALETTI2008,J_Kurths_PhysRep2023,Arenas2008,TANG2014184,Rodrigues2016}.
\\[2mm]
By solving numerically the Kuramoto model, we investigate the effects of damage on networks to identify links where antifragility emerges. We introduce a mathematical framework to assess the impact of damage, defined as a reduction in the weight of a specific link, on identical oscillator systems coupled through weighted networks. Our analysis focuses on how weight modifications in a single link influence global synchronization when the system starts from random initial phases. The measures explored allow the classification of links based on their response to damage, distinguishing those that exhibit antifragility, those where damage is detrimental, and those where it has no significant effect. The paper is organized as follows: in Section \ref{Sec_Theory}, we introduce general definitions and a method to measure the response to damage of coupled oscillators on networks. We also present the Kuramoto model of identical oscillators and its linear
approximation. In Section \ref{Sec_Results}, we apply the methods developed to the study of coupled systems that evolve with the Kuramoto model. We explore different network topologies. Our analysis includes lollipop graphs of varying sizes and a comprehensive evaluation of all the edges of 109 non-isomorphic graphs with six nodes. In
Section \ref{Sec_Conclusions}, we present the conclusions. The approach introduced is general and can be applied to study antifragility in other oscillator systems with different coupling mechanisms, paving the way for the quantitative exploration of antifragility across diverse fields.

\section{General theory}
\label{Sec_Theory}
\subsection{Kuramoto model of identical oscillators}
Let us consider a coupled system of oscillators on a connected, undirected network with $N$ nodes, indexed by $i = 1, \ldots, N$. The phases $\theta_i(t)$ at each node evolve continuously over time $t$, starting at $t = 0$. The network structure is characterized by a symmetric adjacency matrix $\mathbf{A}$, whose elements are defined as $A_{ij} = A_{ji} = 1$ if there is a link connecting nodes $i$ and $j$, and $A_{ij} = A_{ji} = 0$ otherwise. We use the notation $i \to j$ to represent a directed edge starting from node $i$ and ending at node $j$. The links are further described by an $N \times N$ matrix of weights $\mathbf{\Omega}$, with elements $\Omega_{ij} \geq 0$ for $i, j = 1, \ldots, N$. The matrix $\mathbf{\Omega}$ is general in the sense that it incorporates both the network connectivity and specific weights for the links. Based on this information, we define the coupling matrix $\mathbf{W}(\mathbf{\Omega})$, with elements given by
\begin{equation}\label{wijOmega}
	w_{i\to j}(\mathbf{\Omega})\equiv \frac{\Omega_{ij}}{\mathcal{S}_i},
\end{equation}
where $\mathcal{S}_i = \sum_{l=1}^N \Omega_{il}$ represents the generalized strength of node $i$ \cite{ReviewJCN_2021}. In this framework, the entries of the coupling matrix satisfy $0 \leq w_{i \to j}(\mathbf{\Omega}) \leq 1$ and are normalized such that $\sum_{l=1}^N w_{i \to l}(\mathbf{\Omega}) = 1$.
\\[2mm]
The evolution of the phases $\theta_i(t)$ of identical Kuramoto oscillators located at the nodes is governed by a system of coupled nonlinear differential equations
\begin{equation}\label{kuramoto_wij}
	\frac{d\theta_i(t)}{dt}=\sum_{j=1}^{N
	}w_{i\to j}(\mathbf{\Omega})\sin[\theta_j(t)-\theta_i(t)]
\end{equation}
for $i = 1, 2, \ldots, N$. This dynamics represents a generalization of the model introduced by Y. Kuramoto in Ref. \cite{Yoshiki_K} to the case of identical oscillators on weighted networks with a coupling matrix $\mathbf{W}(\mathbf{\Omega})$. For a detailed discussion of the Kuramoto model on networks, we refer the reader to Refs. \cite{Arenas2008, Rodrigues2016, J_Kurths_PhysRep2023}.
\\[2mm]
Furthermore, one of the key features of the system described by Eq. (\ref{kuramoto_wij}) is that the phases can evolve to achieve a globally synchronized state. The conditions under which a system of identical Kuramoto oscillators reaches complete synchronization remain an active area of research. However, the topology of the network governing the coupling is known to play a crucial role \cite{Taylor_2012, Townsend, Ling, HA20101692}. A commonly used quantity to measure the phase coherence of the oscillators is the macroscopic order parameter $r(t)$, defined as \cite{Arenas2008}
\begin{equation}\label{orderparam}
	r(t)=\frac{1}{N}\left|\sum_{j=1}^{N}  \exp\left[\mathbf{i}\,\theta_j(t)\right]\right|,
\end{equation}
where $\mathbf{i} = \sqrt{-1}$. From the definition in Eq. (\ref{orderparam}), it follows that $0 \leq r(t) \leq 1$. In the case of complete phase coherence, $r(t) = 1$, whereas $r(t) = 0$ corresponds to completely incoherent oscillators.
\\[2mm]
The evolution to a completely synchronized state in systems of identical Kuramoto oscillators implies that all oscillators eventually acquire identical phases. Strictly speaking, the time required for these systems to reach a fully synchronized state is infinite. However, this condition can be relaxed \cite{Eraso-Hernandez_2023,Almendral_2007}. For practical purposes, we analyze the time at which the systems are nearly synchronized, defined as the moment when the order parameter $r(t)$ reaches a predetermined fixed value $r$ close to 1. We refer to this specific time as the synchronization time. Equation (\ref{kuramoto_wij}) suggests that synchronization depends on the topology of the network, the coupling matrix $\mathbf{W}(\mathbf{\Omega})$, and the initial conditions of the system $\theta_i(0)$. When the initial phases are randomly chosen with a uniform distribution in the interval $[0, 2\pi)$ for each node, synchronization times can be treated as a stochastic variable. A statistical analysis of this quantity provides valuable insights into the behavior of the system \cite{Eraso-Hernandez_2023}.
\subsection{Functionality}
Previously, we introduced the Kuramoto model of identical oscillators on a weighted network. Let us now consider modifications to the matrix of weights $\mathbf{\Omega}$ and propose a measure to quantify the impact of these changes on global synchronization. Specifically, we examine how damage alters the matrix $\mathbf{\Omega}$, thereby affecting the coupling matrix $\mathbf{W}(\mathbf{\Omega})$ between the oscillators in the network and resulting in variations in synchronization times.
\\[2mm]
In the following, we assume that the function of the system with weights $\mathbf{\Omega}=\mathbf{A}$ and couplings described by $\mathbf{W}(\mathbf{A})$ is to achieve global synchronization starting from random initial phases. To evaluate the synchronization capacity of networks with modified weights, we define a second process incorporating a reorganization of the couplings in $\mathbf{W}(\mathbf{A})$, caused by a reduction in the capacity (damage) of the link $a \to b$ connecting nodes $a$ and $b$ in the network. This modification is represented by the matrix of weights $\mathbf{\Omega}^\star$, whose elements are defined as $\Omega^\star_{ab} = (1 - \beta) A_{ab}$ for the damaged link, and $\Omega^\star_{ij} = A_{ij}$ otherwise. As a consequence of the definition in Eq. (\ref{wijOmega}), the coupling matrices $\mathbf{W}(\mathbf{A})$ and $\mathbf{W}(\mathbf{\Omega}^\star)$ with elements
\begin{equation}\label{w_ij_Omega_explicit}
	w_{i\to j}(\mathbf{\Omega}^\star)\equiv  \frac{\Omega^\star_{ij}}{\sum_{l=1}^N \Omega^\star_{il}}=\cases{
	\displaystyle
	w_{i\to j}(\mathbf{A})\qquad \mathrm{if}\qquad i\neq a,\\ \displaystyle
	\frac{A_{aj}}{k_a-\beta} \hspace{12mm} \mathrm{if}\qquad i=a,\,\, j\neq b,\\
	\displaystyle
	\frac{(1-\beta)A_{ab}}{k_a-\beta} \hspace{5mm} \mathrm{if}\qquad i=a,\,\, j=b,}
\end{equation}
differ in the $a$-th row. The limit $\beta\to 0$ recovers the process used as reference defined by $\mathbf{W}(\mathbf{A})$. In this manner, the matrix of weights codifies the original capacity of the links, motivating the choice $\mathbf{\Omega}=\mathbf{A}$. However, in the modified matrix $\mathbf{\Omega}^\star$, the contribution of the specific link $a \to b$ to the system is reduced by a factor of $(1-\beta)$, representing damage to this particular link. Consequently, the normalization in the coupling matrix $\mathbf{W}(\mathbf{\Omega}^\star)$ redistributes the shortages produced in link $a \to b$ equally among all links connected to node $a$. This type of normalized coupling is relevant in systems where, if a part of the structure fails (in this case, link $a \to b$), the remaining connections must compensate for the loss to maintain the global operation of the system.
\\[2mm]
On the other hand, let $\tau_0$ denote the synchronization time of the original (reference) network with couplings $\mathbf{W}(\mathbf{A})$, and let $\tau(\beta)$ represent the synchronization time when the weights have been altered due to damage, with couplings now described by $\mathbf{W}(\mathbf{\Omega}^\star)$. In both cases, the initial phases at $t = 0$ are chosen randomly with a uniform distribution in the interval $[0,2\pi)$ and maintained identical when evaluating $\tau_0$ and $\tau(\beta)$. Using this information, it is introduced the quantity \cite{Eraso-Hernandez_2023}
\begin{equation}\label{F_beta_edge}
	\mathcal{F}_\beta\equiv\frac{\tau_0}{\tau(\beta)},
\end{equation}
result obtained for one realization of the random initial conditions of the phases. In addition, we refer as ``{\it global functionality}'' to the ensemble average $\left\langle \mathcal{F}_\beta\right\rangle$ for several realizations of the values in Eq. (\ref{F_beta_edge}). The average $\left\langle \mathcal{F}_\beta \right\rangle$ quantifies the global response of the system to a modification with value $\beta$ in the edge $a \to b$. In particular, we have $\left\langle \mathcal{F}_{\beta = 0} \right\rangle = 1$, and when considering the effect of an infinitesimal reduction in the capacity of the link, it is convenient to define
\begin{equation}\label{Lambda_edge}
	\Lambda^{(\mathrm{KM})}\equiv\frac{d\left\langle 	\mathcal{F}_\beta\right\rangle}{d\beta}\Bigg|_{\beta\to 0} .
\end{equation}
The values $\Lambda^{(\mathrm{KM})} > 0$ indicate cases where the reduction of the weight in the link $a \to b$ results in a global response that improves the system's capacity to reach synchronization. This type of response is referred to as antifragile. Conversely, $\Lambda^{(\mathrm{KM})} < 0$ is associated with fragility and reflects the reduction in synchronization capacity caused by the infinitesimal damage in the link $a \to b$. The result $\Lambda^{(\mathrm{KM})} = 0$ is obtained in cases where the infinitesimal change in the weight of a specific link does not affect $\tau(\beta)$. In this regard, these interpretations depend on how the functionality in Eq.~(\ref{F_beta_edge}) is defined. In particular, this measure requires that both systems reach a fixed value $r$ close to 1 within a finite time. However, certain graph structures, such as cyclic graphs, rings, and circulant networks, do not satisfy this condition.  In this sense, our measures in Eqs.~(\ref{F_beta_edge}) and (\ref{Lambda_edge}) are only valid when the analyzed system achieves synchronization in both the original process defined by $\mathbf{\Omega}$, and the modified process given by $\mathbf{\Omega}^\star$.
\\[2mm]
Finally, it is important to highlight that the antifragility measured by Eq.~(\ref{Lambda_edge}) emerges from two combined effects: the reduction of the weight of the connection $a \to b$ and the redistribution of this weight among the links connected to node $a$, due to the normalization effect described in Eq.~(\ref{w_ij_Omega_explicit}). However, the phenomenon of antifragility in synchronization processes can also be observed in systems where the couplings are not normalized (see Section \ref{Section_KM_NNC} for an example illustrating this type of response to damage).
\subsection{Linear approximation of the Kuramoto model}
In the Kuramoto model, it has been shown that for initial conditions closer to synchronization, a linear approximation is valid \cite{Almendral_2007, Grabow_Hill_Grosskinsky_Timme_2010, Grabow_Grosskinsky_Timme_2011}. In Eq. (\ref{kuramoto_wij}), for small values of $\theta_j(t) - \theta_i(t)$, the dynamics of the oscillators can be approximated linearly as
\begin{equation}
	\frac{d\theta_i(t)}{dt}\approx\sum_{j=1}^{N
	}w_{i\to j}(\mathbf{\Omega})[\theta_j(t)-\theta_i(t)]=-\sum_{j=1}^N (\delta_{ij}-w_{i\to j}(\mathbf{\Omega}))\theta_j(t),
	\label{linear_approx1}
\end{equation}
where $\delta_{ij}$ denotes the Kronecker delta. Thus, considering the form of the normalized Laplacian matrix of a weighted network \cite{FractionalBook2019,RiascosDiffusion2023}, we define the elements $\mathcal{L}_{ij}(\mathbf{\Omega})$ of the normalized Laplacian matrix $\hat{\mathcal{L}}(\mathbf{\Omega})$ as
\begin{equation}
	\mathcal{L}_{ij}(\mathbf{\Omega})\equiv \delta_{ij}-w_{i\to j}(\mathbf{\Omega}).
\end{equation}
%
Therefore, the linear approximation in Eq. (\ref{linear_approx1}) defines the dynamical process 
\begin{equation}\label{dtheta_linear}
	\frac{d\theta_i(t)}{dt}=-\sum_{j=1}^N \mathcal{L}_{ij}(\mathbf{\Omega})\theta_j(t).
\end{equation}
The integration of Eq. (\ref{dtheta_linear}) leads to
\begin{equation}\label{linear_exp}
	\theta_i(t)=\sum_{j=1}^N\left(e^{-t \hat{\mathcal{L}}(\mathbf{\Omega})}\right)_{ij}\theta_j(0).
\end{equation}
In addition, using Dirac's notation for the eigenvectors, we have a set of right eigenvectors $\{\left|\phi_j(\mathbf{\Omega})\right\rangle\}_{j=1}^N$ that satisfy the eigenvalue equation
$\hat{\mathcal{L}}(\mathbf{\Omega}) \left|\phi_j(\mathbf{\Omega})\right\rangle = \xi_j(\mathbf{\Omega}) \left|\phi_j(\mathbf{\Omega})\right\rangle$ for $j = 1, \ldots, N$. The eigenvalues $\xi_j(\mathbf{\Omega})$ are ordered by their real part. For connected networks, we have $\xi_1(\mathbf{\Omega}) = 0$ and $0 < \Re\{\xi_2(\mathbf{\Omega})\} \leq \Re\{\xi_3(\mathbf{\Omega})\} \leq \ldots \leq \Re\{\xi_N(\mathbf{\Omega})\}$, where $\Re\{\ldots\}$ denotes the real part of a complex number. Using the set of right eigenvectors, we define the matrix $\mathbf{Q}(\mathbf{\Omega})$ with elements $Q_{ij}(\mathbf{\Omega}) \equiv \left\langle i | \phi_j(\mathbf{\Omega}) \right\rangle$, and the diagonal matrix $\mathbf{\Upsilon}(t, \mathbf{\Omega}) \equiv \textrm{diag}(e^{-t \xi_1(\mathbf{\Omega})}, e^{-t \xi_2(\mathbf{\Omega})}, \ldots, e^{-t \xi_N(\mathbf{\Omega})})$. These matrices satisfy
\begin{equation}
	e^{-t \hat{\mathcal{L}}(\mathbf{\Omega})}=\mathbf{Q}(\mathbf{\Omega})\mathbf{\Upsilon}(t,\mathbf{\Omega})\mathbf{Q}(\mathbf{\Omega})^{-1},
\end{equation}
where $\mathbf{Q}(\mathbf{\Omega})^{-1}$ is the inverse of $\mathbf{Q}(\mathbf{\Omega})$. Using the matrix $\mathbf{Q}(\mathbf{\Omega})^{-1}$, we define the set of left eigenvectors $\{\left\langle \bar{\phi}_i(\mathbf{\Omega})\right|\}_{i=1}^N$ with components $\left\langle \bar{\phi}_i(\mathbf{\Omega})|j\right\rangle=(\mathbf{Q}(\mathbf{\Omega})^{-1})_{ij}$. Therefore, the solution for the linear dynamics in  Eq. (\ref{linear_exp}) takes the form
\begin{equation}\label{theta_i_eigensys}
	\theta_i(t)=\sum_{j=1}^N\sum_{\ell=1}^N e^{-t\xi_\ell(\mathbf{\Omega})}\langle i|\phi_\ell(\mathbf{\Omega})\rangle \langle \bar{\phi}_\ell(\mathbf{\Omega})|j\rangle \theta_j(0).
\end{equation}
Then, the effect of modifications in the matrix of weights $\mathbf{\Omega}$ in the linear approximation defines a process that can be explored analytically using the same methods developed in the study of the Kuramoto dynamics; however, all the temporal evolution is determined by the eigenvalues and eigenvectors of $\hat{\mathcal{L}}(\mathbf{\Omega})$. It is important to note that the linear dynamics is equivalent to the diffusive transport process associated with continuous-time random walks, defined by the normalized Laplacian with elements $\mathcal{L}_{ij}(\mathbf{\Omega})$ (although in our problem, the phase values are not restricted to be probabilities). This equivalence is our main motivation for incorporating the couplings $w_{i \to j}(\mathbf{\Omega})$ given by Eq. (\ref{wijOmega}) into Eq. (\ref{kuramoto_wij}), which define the transition probabilities from node $i$ to $j$ of a random walker on a weighted network (see Ref. \cite{ReviewJCN_2021} for details).
\\[2mm]
The linear approximation in Eq. (\ref{dtheta_linear}) and the analytical solution in Eq. (\ref{theta_i_eigensys}) suggest two additional ways to define the global functionality of a system of coupled oscillators and the effect of infinitesimal modifications. For example, by solving numerically the set of linear equations in Eq. (\ref{dtheta_linear}), we can calculate the synchronization times $\tau^{(\mathrm{linear})}_0$ for the reference process without damage and couplings $\mathbf{W}(\mathbf{A})$, and a time $\tau^{(\mathrm{linear})}(\beta)$ that characterizes the linear dynamics with couplings $\mathbf{W}(\mathbf{\Omega}^\star)$, associated with modifications to the weights in the particular link $a \to b$. In this way, for a given synchronization threshold $r$, and using the same initial random conditions for both processes, we have
\begin{equation}\label{F_beta_edge_linear}
	\mathcal{F}^{(\mathrm{linear})}_\beta\equiv\frac{\tau^{(\mathrm{linear})}_0}{\tau^{(\mathrm{linear})}(\beta)}.
\end{equation}
Therefore, the average over several realizations gives the global functionality $\left\langle \mathcal{F}^{(\mathrm{linear})}_\beta\right\rangle$. Using this quantity, we define the effect of infinitesimal modifications in the link $a \to b$ as
\begin{equation}\label{Lambda_edge_linear}
	\Lambda^{(\mathrm{linear})}\equiv\frac{d\left\langle 	\mathcal{F}^{(\mathrm{linear})}_\beta\right\rangle}{d\beta}\Bigg|_{\beta\to 0}.
\end{equation}
The numerical evaluation of Eqs. (\ref{F_beta_edge_linear}) and (\ref{Lambda_edge_linear}) requires solving Eq. (\ref{dtheta_linear}) or using all the eigenvalues and eigenvectors of $\hat{\mathcal{L}}(\mathbf{A})$ and $\hat{\mathcal{L}}(\mathbf{\Omega}^\star)$. Both approaches lead to the same results but come with high computational costs. An alternative way to reduce the computational cost is to implement a different approach, commonly used in the study of synchronization, by considering the real part of the second eigenvalue $\xi_2(\mathbf{A})$ and $\xi_2(\mathbf{\Omega}^\star)$ of the matrices $\hat{\mathcal{L}}(\mathbf{A})$ and $\hat{\mathcal{L}}(\mathbf{\Omega}^\star)$ \cite{Aguirre_PRL_2014}, respectively.
\\[2mm]
According to Eq. (\ref{theta_i_eigensys}), $\tau^{(\xi_2)}_0 \equiv 1/\Re\{\xi_2(\mathbf{A})\}$ characterizes  the synchronization time in the reference system, whereas $\tau^{(\xi_2)}(\beta) = 1/\Re\{\xi_2(\mathbf{\Omega^\star})\}$  describes the dynamics with modifications in the link $a \to b$. Therefore, it is reasonable to define a global functionality as
\begin{equation}\label{F_beta_edge_second_eig}
	\mathcal{F}^{(\xi_2)}_\beta\equiv\frac{\tau^{(\xi_2)}_0}{\tau^{(\xi_2)}(\beta)}=\frac{\Re\{\xi_2(\mathbf{\Omega^\star})\}}{\xi_2(\mathbf{A})},
\end{equation}
where we use the fact that in undirected connected networks, $\xi_2(\mathbf{A})$ is a real number.
\\[2mm]
A particular feature of the global functionality $\mathcal{F}^{(\xi_2)}_\beta$ is that it does not require averaging over initial conditions, making it easier to obtain than the values $\left\langle \mathcal{F}_\beta\right\rangle$ and $\left\langle \mathcal{F}^{(\mathrm{linear})}_\beta\right\rangle$. Finally, we can evaluate the effects of infinitesimal variations in the link $a \to b$ using
\begin{equation}\label{Lambda_edge_second_eig}
	\Lambda^{(\xi_2)}\equiv\frac{d	\mathcal{F}^{(\xi_2)}_\beta}{d\beta}\Big|_{\beta\to 0}.
\end{equation}
\section{Results}
\label{Sec_Results}
In this section, we present numerical results to illustrate the quantities introduced in Section \ref{Sec_Theory}. The Kuramoto model for identical oscillators is solved numerically using Euler's method (first-order Runge-Kutta) with a time step of $\Delta t=0.001$. It is worth noting that numerical solutions of the Kuramoto model obtained using Runge-Kutta algorithms produce consistent results regardless of the method's order \cite{Muller, Odor2, Bohle}. Additionally, a graphics processing unit (GPU) is employed to parallelize computations, allowing the simultaneous evolution of oscillator systems with varying initial conditions to compute ensemble averages. The parallelization was implemented in python using the open-source JIT compiler Numba \cite{lam2015numba}. Synchronization times are determined using a threshold value $r=0.99$. In Section \ref{Sec_Lollipop}, we investigate the effect of modifications in a lollipop graph with $N=6$. In Section \ref{Sec_lollipop_N} we explore antifragility in lollipops with sizes ranging from $N=4$ to $N=20$.  In Section \ref{Sec_N6} is analyzed the response to infinitesimal modifications in the edges of non-isomorphic graphs with $N=6$ nodes. Finally, in Section~\ref{Section_KM_NNC}, we explore synchronization processes governed by the Kuramoto model in which the couplings are directly defined by the matrix of  weights $\mathbf{\Omega}$.
\subsection{A lollipop graph with $N=6$} 
\label{Sec_Lollipop}
\begin{figure}[t!]
	\centering
	\includegraphics*[width=1.0\textwidth]{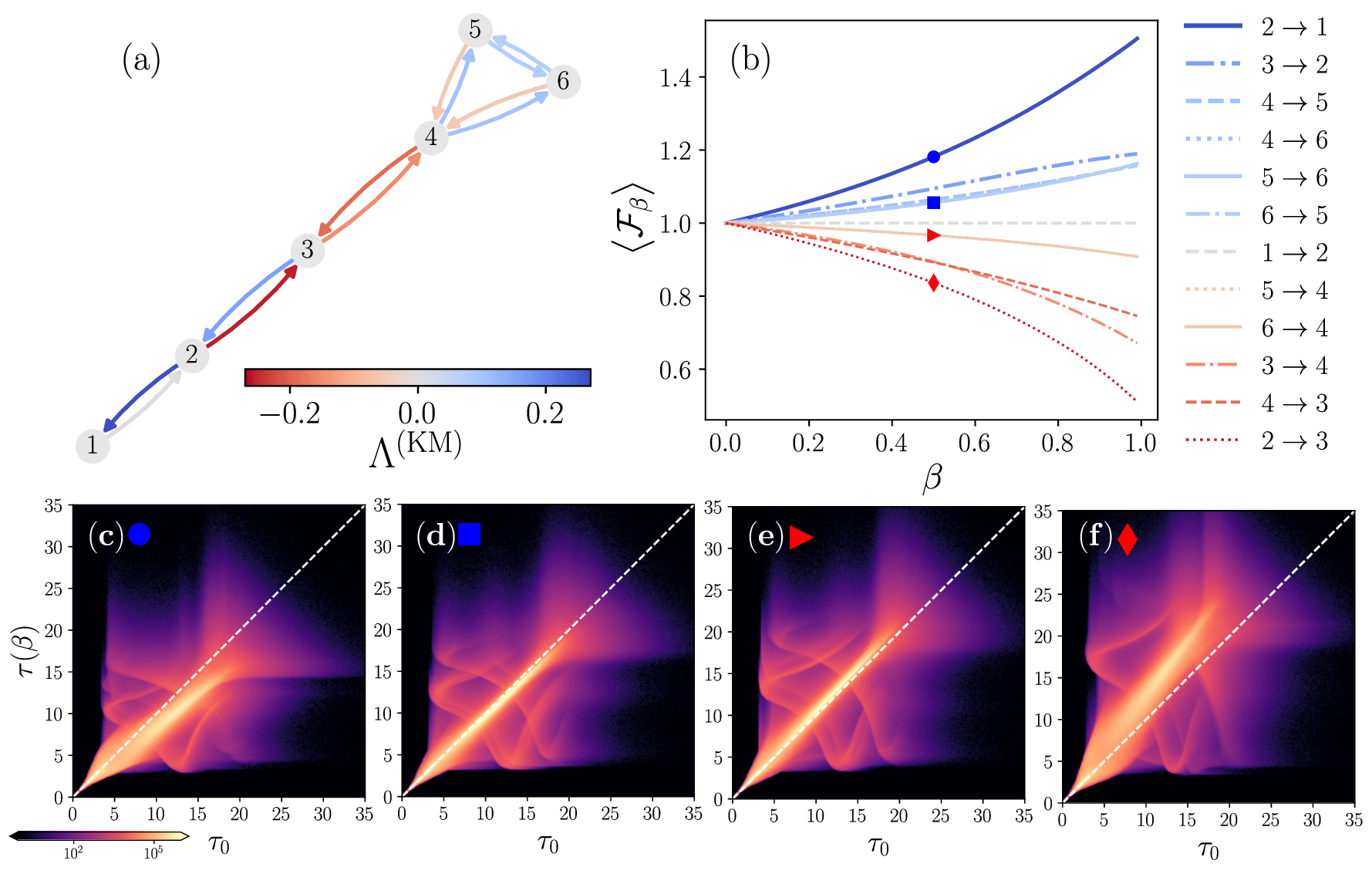}
	\vspace{-4mm}
	\caption{Antifragile response in a lollipop graph with $N=6$ nodes. (a) Representation of the analyzed graph with edges colored according to the value of $\Lambda^{(\mathrm{KM})}$.
		(b) Ensemble average of the functionality $\langle \mathcal{F}_\beta \rangle$ as a function of $\beta$ for all edges in the graph shown in (a). Panels (c) to (f) present two-dimensional histograms of the pairs $(\tau_0, \tau(\beta))$ for $\beta = 0.5$, considering modifications to specific edges: (c) $2 \to 1$, (d) $5 \to 6$, (e) $6 \to 4$, and (f) $2 \to 3$. Dashed lines represent the relation $\tau(\beta)=\tau_0$.  See the main text for details.}
	\label{Fig_1}
\end{figure}
In this section, we study a class of graphs known as lollipops, focusing on cases where the modification of a specific link induces an antifragile response at the global level. A lollipop graph of size $N$ is a network consisting of a fully connected subgraph (clique) with $N-L$ vertices and a linear path graph with $L$ nodes, connected by an additional edge referred to as the bridge. Our main motivation for studying lollipop graphs is that this structure combines a densely connected region with redundant couplings and a linear graph, which can be more fragile due to its lower connection density. In particular, in the context of diffusive transport on networks, cliques and communities exhibit an antifragile response \cite{Eraso_PRE_2024}. With this idea in mind, in the following, we analyze the global functionality $\langle \mathcal{F}_\beta\rangle$ in Eq.~(\ref{F_beta_edge}) and $\Lambda^{(\mathrm{KM})}$ in Eq.~(\ref{Lambda_edge}) to explore the impact of damage in different lollipop graphs.
\\[2mm]
In Fig. \ref{Fig_1}, we analyze a lollipop graph with $N=6$ and $L=3$, as shown in Fig. \ref{Fig_1}(a). Figure  \ref{Fig_1}(b) presents the ensemble average $\langle \mathcal{F}_\beta \rangle$ as a function of $\beta \in [0,1)$ for all directed edges in the graph. We use the notation $i \to j$ in the labels to indicate the direction of the edge, starting from node $i$ and ending at node $j$. The results were obtained using $10^6$ realizations to compute each value of $\langle \mathcal{F}_\beta \rangle$. Additionally, Fig. \ref{Fig_1}(a) includes the values of $\Lambda^{(\mathrm{KM})}$ for each edge, encoded in the colorbar in Fig. \ref{Fig_1}(a). These values were calculated with the derivative approximation in Eq. (\ref{Lambda_edge}) 
\begin{equation}\label{Lambda_edge_numerical}
	\Lambda^{(\mathrm{KM})}\approx\frac{1}{\Delta \beta}\left[\langle 	\mathcal{F}_{\Delta\beta}\rangle-1\right]
\end{equation}
with $\Delta \beta=0.01$ and using $10^{7}$ realizations to evaluate $\langle 	\mathcal{F}_{\Delta\beta}\rangle$ in each edge. The curves in Fig. \ref{Fig_1}(b) are also colored according to the corresponding $\Lambda^{(\mathrm{KM})}$ values.
\\[2mm]
In Figs. \ref{Fig_1}(c)-(f), we present the statistical analysis of the synchronization times $\tau_0$ and $ \tau(\beta)$  in four particular cases considering modifications with $\beta=0.5$ in different edges of the graph. Each histogram is obtained from the statistical analysis of synchronization using $10^9$ realizations of identical oscillators, with random initial phases uniformly distributed in the interval $[0, 2\pi)$. For the calculation of each pair $(\tau_0, \tau(\beta))$, the same initial conditions are maintained. The results show the effect of modifications in the links $2 \to 1$ [Fig. \ref{Fig_1}(c)], $5 \to 6$ [Fig. \ref{Fig_1}(d)], $6 \to 4$ [Fig. \ref{Fig_1}(e)], and $2 \to 3$ [Fig. \ref{Fig_1}(f)]. 
\\[2mm]
The results in Fig. \ref{Fig_1} demonstrate the system's antifragile or fragile global response to reductions in edge weights, depending on the specific edge modified. The values of $\Lambda^{(\mathrm{KM})}$ shown in Fig. \ref{Fig_1}(a), which correspond to infinitesimal edge modifications, allow the identification of antifragile links where $\Lambda^{(\mathrm{KM})} > 0$. For these links, a reduction in the weight enhances the system's synchronization capacity, resulting in $\langle \mathcal{F}_\beta \rangle > 1$. This indicates that, for a significant fraction of initial conditions, $\tau(\beta) < \tau_0$.  Similarly, for links with $\Lambda^{(\mathrm{KM})} < 0$, a reduction in the weight is detrimental, leading to a global fragile response characterized by $\langle \mathcal{F}_\beta \rangle < 1$. Additionally, in the case where $\Lambda^{(\mathrm{KM})} = 0$ for the link $1 \to 2$, changes with $\beta$ preserve the couplings $\mathbf{W}(\mathbf{\Omega}^\star)$ due to the normalization imposed by Eq. (\ref{w_ij_Omega_explicit}). The different global responses of the oscillators, identified through infinitesimal variations in Eq. (\ref{Lambda_edge_numerical}), remain valid for the interval $\beta \in (0,1)$, as shown in Fig. \ref{Fig_1}(b). For the analyzed lollipop graph, links with $\Lambda^{(\mathrm{KM})} > 0$ exhibit a monotonic increase in $\langle \mathcal{F}_\beta \rangle$, whereas links with $\Lambda^{(\mathrm{KM})} < 0$ show a monotonic decrease in functionality across the entire interval $0<\beta <1$.
\\[2mm]
Finally, it is important to emphasize that the values of $\langle \mathcal{F}_\beta \rangle$ provide only average information about the response to changes in synchronization times. Specifically, they reflect the differences between the average synchronization time $\tau_0$ of the oscillators in the original system and $\tau(\beta)$ in the modified system. In Figs. \ref{Fig_1}(c)-(f), we extend the analysis by constructing two-dimensional histograms of the pairs $(\tau_0, \tau(\beta))$ for $\beta = 0.5$, based on $10^9$ realizations of the initial conditions of the oscillator phases. The corresponding values of $\langle \mathcal{F}_\beta \rangle$ are highlighted with markers in Fig. \ref{Fig_1}(b), where two cases exhibiting an antifragile response and two cases with a fragile response are considered.
\\[2mm]
Figure \ref{Fig_1}(c) illustrates a case where the weight reduction occurs on the link $2 \to 1$, which has the highest value of $\Lambda^{(\mathrm{KM})}$ and the largest increase in $\langle \mathcal{F}_\beta \rangle$. The results for this case indicate that, for a significant number of initial conditions, $\tau(\beta) < \tau_0$. This is evidenced by a greater density of points $(\tau_0, \tau(\beta))$ below the dashed line representing $\tau(\beta) = \tau_0$. This imbalance leads to a value of $\langle \mathcal{F}_{\beta=0.5} \rangle = 1.181$. A similar analysis is performed for a change with $\beta = 0.5$ on the link $5 \to 6$, which results in $\langle \mathcal{F}_{\beta=0.5} \rangle = 1.056$ and also exhibits an antifragile response. In contrast, Figs. \ref{Fig_1}(e)-(f) depict cases with a fragile response, where a larger number of initial states produce responses with  $\tau(\beta) > \tau_0$. In particular, the fragile response shown in Fig. \ref{Fig_1}(f), corresponding to modifications in the edge $2 \to 3$, is the most pronounced. This case exhibits the smallest value of $\Lambda^{(\mathrm{KM})}$ and the largest reduction in $\langle \mathcal{F}_\beta \rangle$, with $\langle \mathcal{F}_{\beta=0.5} \rangle = 0.836$.
\begin{figure*}[t!]
	\centering
	\includegraphics*[width=1.0\textwidth]{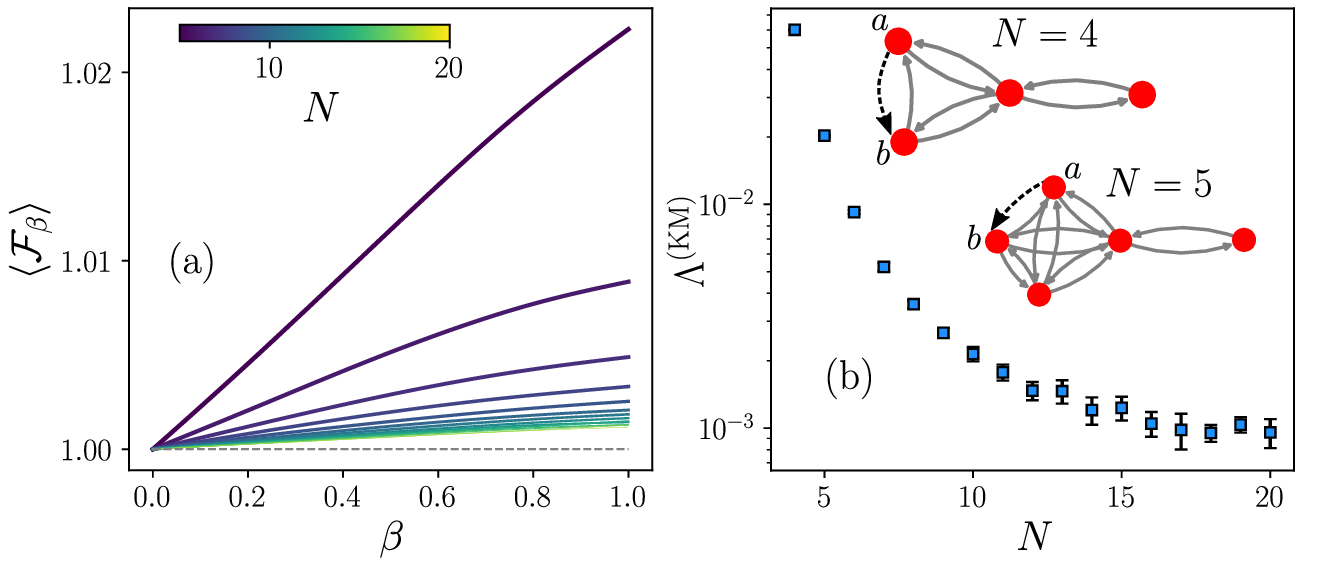}
	\vspace{-5mm}
	\caption{Antifragility in lollipop graphs with $L=1$ and different sizes $N$. (a) Ensemble average of the functionality $\langle \mathcal{F}_\beta \rangle$ as a function of $\beta$ for lollipop graphs. The colorbar encodes the sizes explored, ranging from $5 \leq N \leq 20$. (b) Values of $\Lambda^{(\mathrm{KM})}$ for different graph sizes $N$. In all cases shown in (a) and (b), the analysis focuses on the effect of reducing the weight of a link $a \to b$ within the clique, the chosen link is not directly connected to the node forming part of the bridge. This type of edge is illustrated with a dashed arrow and shown as insets in (b) for the graphs with $N = 4$ and $N = 5$. See the main text for further details. }
	\vspace{2mm}
	\label{Fig_2}
\end{figure*}
\subsection{Antifragility in lollipop graphs}
\label{Sec_lollipop_N}
Our findings reported in Fig. \ref{Fig_1} reveal the antifragile response observed in the link $5 \to 6$ (or $6 \to 5$), which is of particular interest. This type of link belongs to the clique, a structure where transportation processes involving random walks under damage conditions have demonstrated a strong antifragile response to weight reductions (see Ref. \cite{Eraso_PRE_2024} for details). Motivated by this observation, in Fig. \ref{Fig_2}, we analyze the effects of modifications in lollipop graphs with $L=1$ and sizes varying from $N=4$ to $N=20$.  We analyze the effect of reducing the weight of a link $a \to b$ in the clique that is not connected directly with the node that forms part of the bridge [this type of edge is illustrated in the link $a\to b$ in the graphs with $N=4,\, 5$ presented as inset in Fig. \ref{Fig_2}(b)]. 
\\[2mm]
In Fig. \ref{Fig_2}(a), we present the numerical results for $\langle \mathcal{F}_\beta \rangle$ as a function of $\beta$ with $0 \leq \beta < 1$ for graphs with $L=1$ and sizes $5 \leq N \leq 20$, as indicated by the colorbar. These values are obtained numerically using $10^6$ realizations for each $\beta \in [0,1)$, chosen with increments of $\Delta \beta = 0.02$. In all cases, $\langle \mathcal{F}_\beta \rangle > 1$ for $0 < \beta < 1$ (the limit $\langle \mathcal{F}_\beta \rangle=1$ is presented with a dashed line), indicating that the network's capacity of synchronization increases with the effect of $\beta$ in the link $a \to b$. Additionally, the numerical results reveal that the reduction of the weight in this edge has a more pronounced impact on smaller structures.
\\[2mm]
On the other hand, the antifragility observed in Fig. \ref{Fig_2}(a) can be further analyzed by examining the derivative of $\langle \mathcal{F}_\beta \rangle$ evaluated at $\beta = 0$. Specifically, the slope of the tangent line to $\langle \mathcal{F}_\beta \rangle$ at $\beta = 0$ provides the value of $\Lambda^{(\mathrm{KM})}$, which indicates how a small, infinitesimal modification to the link $a \to b$ alters  the global functionality.  To quantify the numerical error of this measure, we analyze $10$ values of $\Lambda^{(\mathrm{KM})}$, each computed numerically using the approximation in Eq. (\ref{Lambda_edge_numerical}) with $10^7$ realizations and $\Delta \beta = 0.01$.  In Fig. \ref{Fig_2}(b), we present $\Lambda^{(\mathrm{KM})}$ as a function of $N$ for lollipop graphs with $L = 1$ and sizes $4 \leq N \leq 20$. Square markers indicate the average values of $\Lambda^{(\mathrm{KM})}$, while the error bars represent the standard deviation of the computed values. The results show that the numerical variations in $\Lambda^{(\mathrm{KM})}$ lead to an error on the order of $10^{-4}$. The results in Fig. \ref{Fig_2}(b) illustrate how the antifragility of the analyzed link diminishes with increasing graph size $N$. Our findings indicate that in all cases studied, the response remains antifragile, with $\Lambda^{(\mathrm{KM})} > 0$. However, the global impact of infinitesimal variations in the weights of this connection decreases significantly as the clique size grows, with a marked decline in graphs of size $N \leq 15$. For $N > 15$, the reduction is less pronounced, but the uncertainties introduced by the numerical method become more noticeable, imposing limitations on the analysis of larger graphs with $N > 20$.
\subsection{Graphs with $N=6$}
\label{Sec_N6}
The method described in Sections \ref{Sec_Lollipop} and  \ref{Sec_lollipop_N} for studying antifragility in lollipop graphs can be applied to analyze the impact of infinitesimal modifications on individual links in a graph. This approach allows the detection of edges whose reduction in their weight enhances, preserves, or diminishes the global capacity of the structure to reach synchronization. In this section, we study $\Lambda^{(\mathrm{KM})}$ for the edges of all connected non-isomorphic graphs with $N = 6$ nodes. The analyzed graph set is available in Ref. \cite{ConnectedGraphs} and includes 112 graphs with several topologies, such as diverse trees (e.g., linear and star graphs), networks with cycles and cliques of different sizes, and structures with high-density of links, including the fully connected graph. However, we focus exclusively on graphs that fully synchronize from random initial conditions. We exclude three specific cases: a ring with six nodes and two networks formed by a ring with five nodes plus one or two additional edges. Consequently, the final dataset analyzed consists of 109 graphs.
\\[2mm]
\begin{figure*}[t!]
	\centering
	\includegraphics*[width=1.0\textwidth]{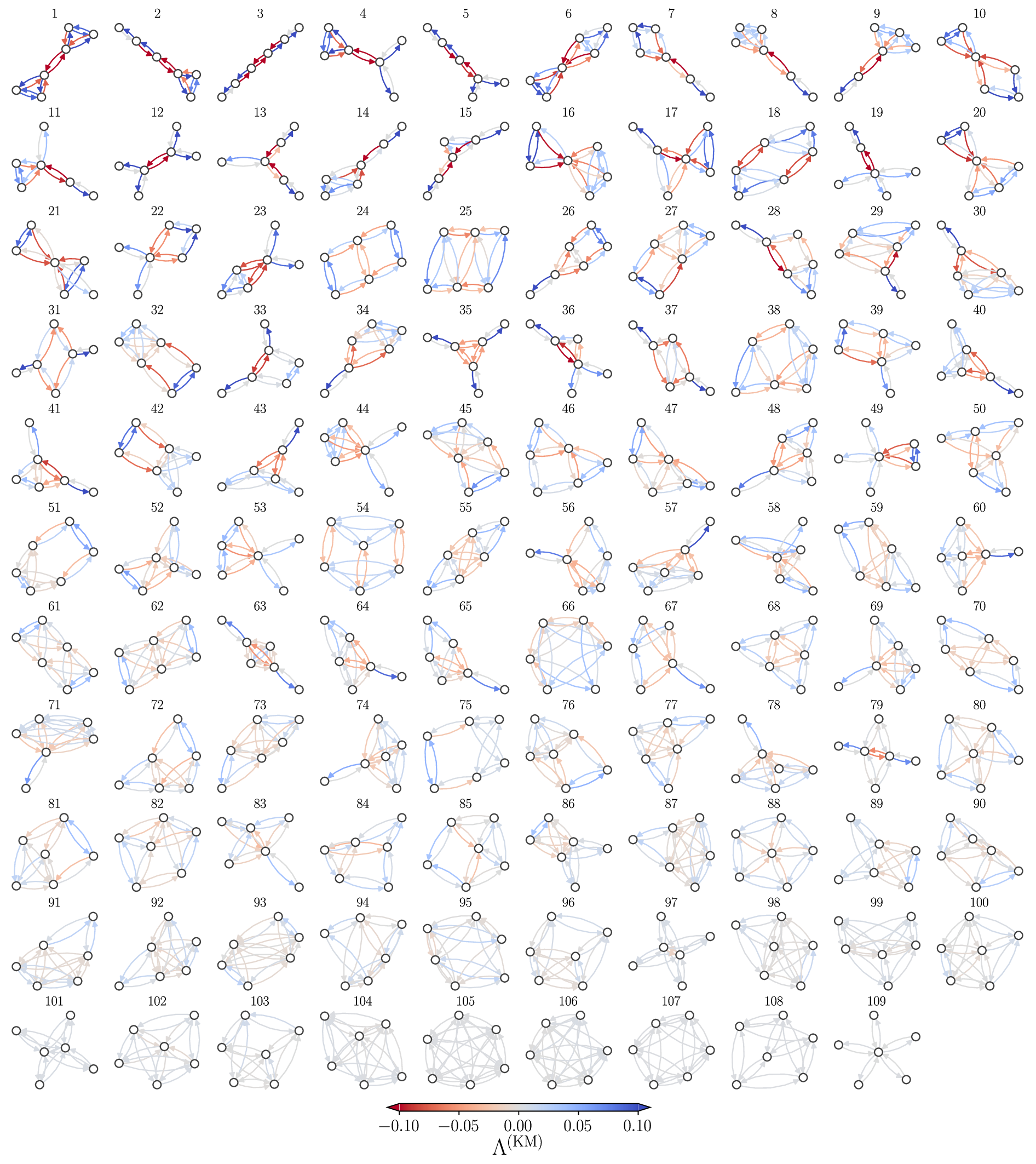}
	\vspace{-5mm}
	\caption{ Response to infinitesimal modifications in the links of non-isomorphic connected graphs with $N = 6$ nodes \cite{ConnectedGraphs}. The values are computed numerically using Eq. (\ref{Lambda_edge_numerical}), considering infinitesimal damage to each link  with $\Delta \beta=0.01$ and $10^7$ realizations. The results for $\Lambda^{\mathrm{(KM)}}$ are represented using the colorbar.   }
	\label{Fig_3}
\end{figure*}
\begin{figure*}[t!]
	\centering
	\includegraphics*[width=1.0\textwidth]{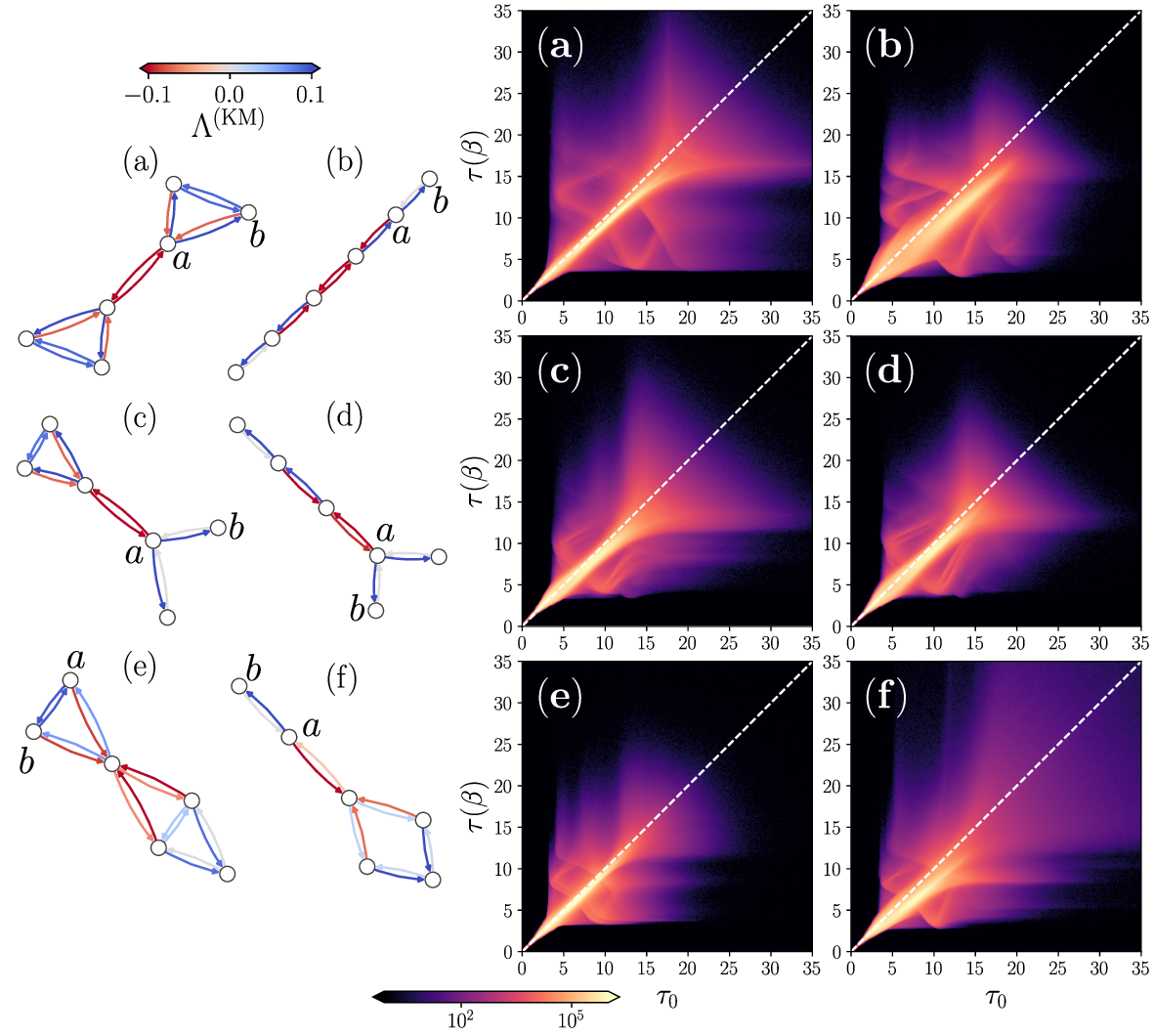}
	\vspace{-5mm}
	\caption{Statistical analysis of the synchronization times $\tau_0$ and $\tau(\beta)$ for different graphs exhibiting an antifragile response. The left panels (a)-(f) show the analyzed structures, corresponding to graphs 1 and $3,\dots,7$ in Fig. \ref{Fig_3}. The colorbar encodes the values of $\Lambda^{\mathrm{(KM)}}$ for all edges, while the letters $a$ and $b$ indicate the nodes of the directed link $a \to b$ with the largest antifragility. The right panels display the corresponding two-dimensional histograms of $(\tau_0, \tau(\beta))$, obtained by modifying the edge $a \to b$ with $\beta = 0.5$.  Dashed lines represent the relation $\tau(\beta)=\tau_0$.  See the main text for details.}
	\label{Fig_4}
\end{figure*}
In Fig. \ref{Fig_3}, we present the numerical results obtained for $\Lambda^{(\mathrm{KM})}$, which evaluate the effect of infinitesimal modifications on each link in 109 non-isomorphic graphs. In this representation, each link is colored according to its respective $\Lambda^{(\mathrm{KM})}$, calculated numerically for each directed edge using Eq. (\ref{Lambda_edge_numerical}) with $\Delta \beta=0.01$ and $10^7$ realizations to obtain the ensemble average, the values are codified in the colorbar. The graphs are sorted using the sum of the values $\Lambda^{(\mathrm{KM})}$ restricted to the antifragile links (those with with $\Lambda^{(\mathrm{KM})}>0$). Our findings show how antifragility emerge in the synchronization of oscillators and reveal that it is a challenge to identify a general rule for the distribution of links with $\Lambda^{(\mathrm{KM})} > 0$ within the graph. However, in the first nine graphs, it is evident that structures connecting linear subgraphs with cliques or highly connected subgraphs exhibit links with greater antifragility. This pattern is observed in graph 1 (a barbell graph) formed by two cliques, the lollipop graphs 2 and 8, the trees 3 and 5, graph 4 (derived from the barbell graph by removing two directed links in one clique), and graph 6, which connects a clique with three nodes with a structure with four nodes. Graphs 7 and 9 similarly connect a subgraph with four nodes densely connected to a linear graph with two nodes.
\\[2mm]
In Fig. \ref{Fig_4}, we complement the analysis of the antifragile response of the first graphs in Fig. \ref{Fig_3} by exploring the synchronization times under modifications of the link $a\to b$ with the highest value of $\Lambda^{(\mathrm{KM})}$. For this link, we compute $10^9$ realizations of the synchronization times $\tau_0$ and $\tau(\beta)$ with $\beta=0.5$, using random initial phases. The analyzed graphs are shown in the left panels of Fig. \ref{Fig_4}, while the corresponding two-dimensional histograms of $(\tau_0, \tau(\beta))$ are presented in the right panels. The numerical values are obtained by modifying the edge $a \to b$, with the respective nodes $a$ and $b$ highlighted in each graph. We examine graphs $1$ and $3$–$7$ from Fig. \ref{Fig_3}, while graph 2 was previously analyzed in detail in Fig. \ref{Fig_1}; in particular, Fig. \ref{Fig_1}(c) presents the results for the link with the highest antifragility. In Table \ref{Tab_1}, we report the ensemble average $\langle \tau_0 \rangle$ of the times $\tau_0$ for the structure without damage, the  highest value of $\Lambda^{(\mathrm{KM})}$ and the response to damage $\langle \mathcal{F}_{\beta=0.5} \rangle$ for the most antifragile link, each ensemble average is obtained using $10^9$ realizations.
\\[2mm]
Our findings in Fig. \ref{Fig_4} and Table \ref{Tab_1} indicate that for the analyzed links, a classification of $\Lambda^{(\mathrm{KM})}>0$ under infinitesimal weight variations consistently corresponds to an antifragile response, characterized by $\langle \mathcal{F}_{\beta} \rangle > 1$ at $\beta=0.5$. Additionally, the two-dimensional histograms of $(\tau_0, \tau(\beta))$ reveal a higher fraction of synchronization times from random initial phases satisfying $ \tau(\beta)<\tau_0 $, meaning that more values fall below the dashed line $\tau_0 = \tau(\beta)$.
\\[2mm]
\begin{table}[t]
	\centering
	\begin{tabular}{c c  c  c }
		
		{\bf Graph} &  $\langle \tau_0 \rangle$ &  $\Lambda^{(\mathrm{KM})}$  & $\langle \mathcal{F}_{\beta=0.5} \rangle$ \\[0.5mm]
		\hline
		1        & 9.3125        &   $1.321\times 10^{-1}$        & 1.0823         \\  
		2         & 9.5614        &  $2.653\times 10^{-1}$        &  1.1815        \\  
		3         & 9.8422        &  $2.258\times 10^{-1}$         &  1.1536        \\  
		4         &  7.5638       &  $1.526\times 10^{-1}$         &   1.0949       \\  
		5         &  8.1089       &  $1.202\times 10^{-1}$         &   1.0756       \\      
		6         &  6.0817       &  $9.364\times 10^{-2}$         &  1.0665        \\
		7         &  7.0777       &  $2.667\times 10^{-1}$         &  1.1754        \\
		\hline
	\end{tabular}
	\caption{\label{Tab_1}Characterization of synchronization and antifragile response in the first seven graphs analyzed in Fig. \ref{Fig_3}.}
\end{table}
\begin{figure*}[t!]
	\centering
	\includegraphics*[width=1.0\textwidth]{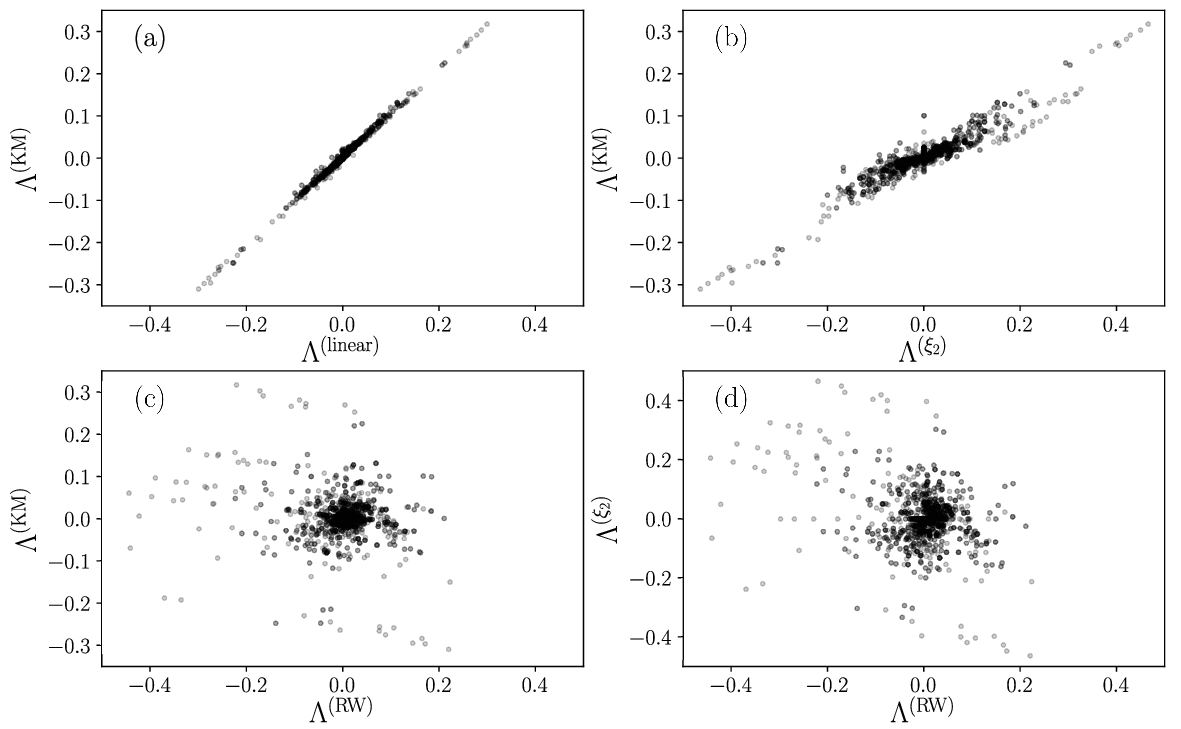}
	\vspace{-5mm}
	\caption{Different metrics describing the effect of infinitesimal variations in the link weights for the 109 graphs analyzed in Fig. \ref{Fig_3}. The panels present diverse combinations of the values $\Lambda^{(\mathrm{KM})}$ in Eq. (\ref{Lambda_edge_numerical}), $\Lambda^{(\mathrm{linear})}$ in Eq. (\ref{Lambda_edge_linear_numerical}), $\Lambda^{(\xi_2)}$ in Eq. (\ref{Lambda_edge_second_eig}) and $\Lambda^{(\mathrm{RW})}$ in Eq. (\ref{Lambda_edge_RW_numerical}). Further details can be found in the main text. }
	\vspace{2mm}
	\label{Fig_5}
\end{figure*}
After calculating $\Lambda^{(\mathrm{KM})}$ for all the edges in the graphs analyzed in Fig. \ref{Fig_3}, we use this information to investigate the relationship between $\Lambda^{(\mathrm{KM})}$ and other quantities introduced in Section \ref{Sec_Theory} for evaluating dynamical systems associated with the Kuramoto model. To this end, in Fig. \ref{Fig_5}, we present scatter plots of points with coordinates $(\Lambda_x, \Lambda_y)$, where $\Lambda_x$ and $\Lambda_y$ represent two measures of the response to modifications of the same edge in the graph. The quantities analyzed include $\Lambda^{(\mathrm{KM})}$ defined in Eq. (\ref{Lambda_edge_numerical}), $\Lambda^{(\mathrm{linear})}$ in Eq. (\ref{Lambda_edge_linear}), $\Lambda^{(\xi_2)}$ in Eq. (\ref{Lambda_edge_second_eig}), and $\Lambda^{(\mathrm{RW})}$ in Eq. (\ref{Lambda_edge_RW_numerical}), which characterizes the effect of damage in transportation processes associated with random walks on networks \cite{Eraso_PRE_2024}, see Appendix \ref{Appendix_RW} for details.
\\[2mm]
In Fig. \ref{Fig_5}(a), we compare $\Lambda^{(\mathrm{KM})}$ with $\Lambda^{(\mathrm{linear})}$. Specifically, calculating $\Lambda^{(\mathrm{linear})}$ involves numerically solving the linear system of equations in Eq. (\ref{dtheta_linear}) to determine the synchronization times $\tau^{(\mathrm{linear})}_0$ and $\tau^{(\mathrm{linear})}(\beta)$ starting from random initial phases. By using these values is computed numerically 
\begin{equation}\label{Lambda_edge_linear_numerical}
	\Lambda^{(\mathrm{linear})}=\frac{1}{\Delta \beta}\left[ \left\langle\mathcal{F}_\beta^{(\mathrm{linear})}\right\rangle-1\right].
\end{equation} 
The results are obtained with $\Delta \beta = 0.01$ and $10^{7}$ realizations. Our findings in Fig. \ref{Fig_5}(a) reveal a linear relationship between $\Lambda^{(\mathrm{KM})}$ and $\Lambda^{(\mathrm{linear})}$, with subtle deviations from this trend. These deviations indicate that, in most cases, the Kuramoto model and its linear approximation yield similar classifications of edges as antifragile, neutral, or fragile in the analyzed graphs. However, an important consideration when evaluating $\Lambda^{(\mathrm{KM})}$ and $\Lambda^{(\mathrm{linear})}$ is that both have the same computational cost, as they require solving differential equations numerically and averaging over initial conditions. Given this, it is more practical to compute $\Lambda^{(\mathrm{KM})}$ to assess the response of links in the network.
\\[2mm]
In contrast, Fig. \ref{Fig_5}(b) presents the relationship between $\Lambda^{(\mathrm{KM})}$ and $\Lambda^{(\xi_2)}$. Calculating $\Lambda^{(\xi_2)}$ is computationally faster than $\Lambda^{(\mathrm{KM})}$, as evaluating Eq. (\ref{Lambda_edge_second_eig}) requires only the second smallest eigenvalue of the matrices $\hat{\mathcal{L}}(\mathbf{A})$ and $\hat{\mathcal{L}}(\mathbf{\Omega}^\star)$. The scatter plot of points in Fig. \ref{Fig_5}(b) reveals a linear relationship between $\Lambda^{(\mathrm{KM})}$ and $\Lambda^{(\xi_2)}$, with notable deviations. In particular, for $|\Lambda^{(\mathrm{KM})}| < 0.1$, several cases exhibit discrepancies in the signs of $\Lambda^{(\mathrm{KM})}$ and $\Lambda^{(\xi_2)}$, leading to different classifications of edge responses to modifications. However, for $|\Lambda^{(\mathrm{KM})}| \geq 0.1$, the classifications of fragility or antifragility are consistent between both measures. These findings suggest that $\Lambda^{(\xi_2)}$ can serve as a fast method to identify edges with significant fragile or antifragile responses to weight modifications.
\\[2mm]
Furthermore, the relationship between the linear dynamics in Eq. (\ref{dtheta_linear}), defined by the modified Laplacian $\hat{\mathcal{L}}(\mathbf{\Omega})$, and the transition matrix of a random walker on a weighted network, $\mathbf{W}(\mathbf{\Omega})$, motivates the exploration of how our findings relate to antifragility when the network's primary function is  to reach efficiently all the nodes through the movement of a random walker. In Fig. \ref{Fig_5}(c), we present the pairs $(\Lambda^{(\mathrm{RW})}, \Lambda^{(\mathrm{KM})})$, where $\Lambda^{(\mathrm{RW})}$ is calculated using the numerical approximation of Eq. (\ref{Lambda_edge_RW}) in Appendix \ref{Appendix_RW}
\begin{equation}\label{Lambda_edge_RW_numerical}
	\Lambda^{(\mathrm{RW})}\approx \frac{1}{\Delta \beta}\left[\mathcal{F}^{(\mathrm{RW})}_{\Delta \beta}-1\right]
\end{equation}
with $\Delta \beta=10^{-4}$. In this case, the distribution of points in Fig. \ref{Fig_5}(c) shows no discernible relationship between $\Lambda^{(\mathrm{RW})}$ and $\Lambda^{(\mathrm{KM})}$. Consequently, there is no direct connection between the antifragile or fragile response measured by these quantities, highlighting that the classification of edges based on their response to damage depends on the graph, the perturbed link, and the specific dynamical process occurring on the structure. As an additional test to the measures explored, Fig. \ref{Fig_5}(d) presents an analysis of the quantities $\Lambda^{(\xi_2)}$ and $\Lambda^{(\mathrm{RW})}$. Once again, no correlation is observed between these two measurements, indicating no relationship between the antifragility measured in diffusive transport processes on networks and the dynamics whose global function is the synchronization of coupled oscillators.
\\[2mm]
\begin{figure*}[t!]
	\centering
	\includegraphics*[width=1.0\textwidth]{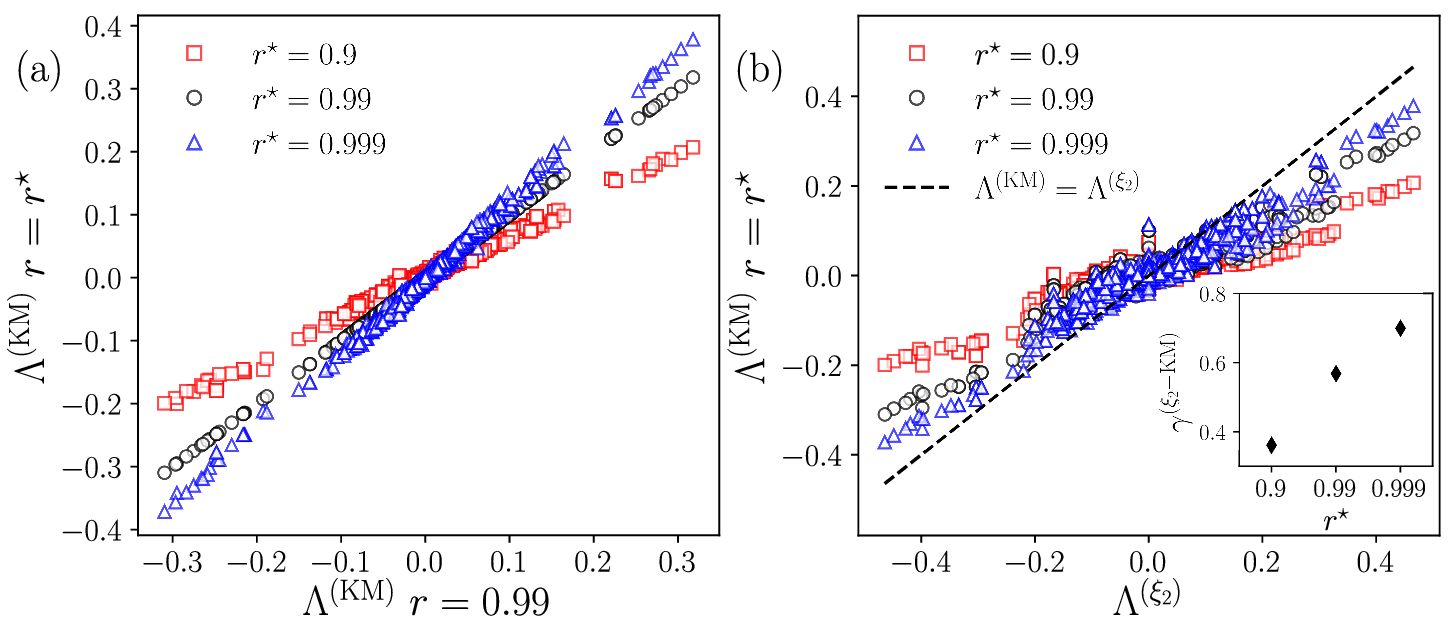}
	\vspace{-5mm}
	\caption{Effect of the threshold $r$ on the numerical values of $\Lambda^{(\mathrm{KM})}$ for all edges of the graphs with $N = 6$ in Fig.~\ref{Fig_3}. (a) Scatter plot of points $(\Lambda_x, \Lambda_y)$, where $\Lambda_x$ corresponds to the values of $\Lambda^{(\mathrm{KM})}$ obtained with $r = 0.99$, and $\Lambda_y$ represents the corresponding values of $\Lambda^{(\mathrm{KM})}$ computed using $r = r^\star$, with $r^\star = 0.9$, $0.99$, and $0.999$.  (b) In this panel, we repeat the same visualization as in (a), but now $\Lambda_x$ denotes the values of $\Lambda^{(\xi_2)}$  defined in Eq.~(\ref{Lambda_edge_second_eig}). The inset depicts the results obtained for the slope $\gamma^{(\xi_2-\mathrm{KM})}$ from the linear fit of each  set of points as a function of the respective $r^\star$. }
	\vspace{2mm}
	\label{Fig_6}
\end{figure*}
Once $\Lambda^{(\mathrm{KM})}$ and its relationship with other related quantities have been explored in Fig.~\ref{Fig_5}, it is important to note that all the results presented for $\Lambda^{(\mathrm{KM})}$ in Figs.~\ref{Fig_3}--\ref{Fig_5} rely on the functionality $\mathcal{F}_\beta$ defined in Eq.~(\ref{F_beta_edge}), where the synchronization times are computed using the threshold $r = 0.99$ for the order parameter in Eq.~(\ref{orderparam}). However, it is crucial to understand how the classification of antifragility and fragility provided by $\Lambda^{(\mathrm{KM})}$ for all edges in the graphs of Fig.~\ref{Fig_3} is affected by the choice of different values of the threshold $r$. To this end, in Fig.~\ref{Fig_6} we analyze the values of $\Lambda^{(\mathrm{KM})}$ obtained using three different thresholds for the order parameter: $r = 0.9$, $0.99$, and $0.999$, considering all edges of the graphs with $N = 6$ explored in Fig.~\ref{Fig_3} with $10^7$ realizations to obtain $\Lambda^{(\mathrm{KM})}$.
\\[2mm]
In Fig.~\ref{Fig_6}(a), we present the pairs $(\Lambda_x, \Lambda_y)$, where $\Lambda_x$ corresponds to the values computed using the threshold $r = 0.99$, these values were previously shown in Fig.~\ref{Fig_3} and analyzed in Fig.~\ref{Fig_5}. The corresponding $\Lambda_y$ denotes $\Lambda^{(\mathrm{KM})}$ obtained for the same links but using the thresholds $r = r^\star$, with $r^\star = 0.9$, $0.99$, and $0.999$. The resulting scatter plot is shown using different markers for each value of $r^\star$. The numerical results reported in Fig.~\ref{Fig_6}(a) reveal a strong linear relationship between the values of $\Lambda^{(\mathrm{KM})}$ computed with $r = 0.99$ and those obtained using the thresholds $r = 0.9$ and $r = 0.999$. Additionally, it is evident that $|\Lambda^{(\mathrm{KM})}|$ tends to increase as $r$ approaches 1. In particular, for links with $\Lambda^{(\mathrm{KM})} > 0$, the antifragility increases for higher values of $r$. It is also worth noting that the antifragility/fragility classifications derived from $\Lambda^{(\mathrm{KM})}$ for the $r$ explored remain similar to those presented in Fig. \ref{Fig_3}.
\\[2mm]
In Fig.~\ref{Fig_6}(b), we complement the analysis by exploring the relationship between the values of $\Lambda^{(\mathrm{KM})}$ obtained for $r = r^\star = 0.9$, $0.99$, and $0.999$, and the corresponding values of $\Lambda^{(\xi_2)}$ calculated using Eq.~(\ref{Lambda_edge_second_eig}). The latter has the advantage of not requiring ensemble averages or the specification of a particular threshold, providing a more direct method to characterize the system's response to damage. The results obtained from the scatter plot of points $(\Lambda^{(\xi_2)}, \Lambda^{(\mathrm{KM})})$ reveal linear trends. However, noticeable deviations suggest that the approximation given by $\Lambda^{(\xi_2)}$ does not fully capture the impact of damage as quantified by $\Lambda^{(\mathrm{KM})}$. Furthermore, by performing a linear fit of the form $\Lambda^{(\mathrm{KM})} = \gamma^{(\xi_2-\mathrm{KM})} \Lambda^{(\xi_2)} + \mathcal{C}$, we find that $\mathcal{C} \approx 0.003$, a value that is effectively within the numerical error margin of $\Lambda^{(\mathrm{KM})}$. The corresponding slopes $\gamma^{(\xi_2-\mathrm{KM})}$ for each case are reported in the inset of Fig.~\ref{Fig_6}(b). It is observed that $\gamma^{(\xi_2-\mathrm{KM})}$ increases with $r$, approaching the limiting case in which $\Lambda^{(\mathrm{KM})} = \Lambda^{(\xi_2)}$. In general, the analysis presented in Fig.~\ref{Fig_6}(b) provides additional evidence of antifragility in graphs and shows that this effect can also be captured through $\Lambda^{(\xi_2)}$, which is computationally more accessible. This alternative measure enables the identification of links with a strong response to damage, particularly when $|\Lambda^{(\mathrm{KM})}| \geq 0.2$. For lower values, i.e., when $|\Lambda^{(\mathrm{KM})}| < 0.2$, nonlinear effects may be important in the evaluation of the consequences of damage. These effects are captured by $\Lambda^{(\mathrm{KM})}$ but not by $\Lambda^{(\xi_2)}$.
\subsection{Antifragility in a system with non-normalized couplings}
\label{Section_KM_NNC}
In the Kuramoto model defined by Eq. (\ref{kuramoto_wij}), the interactions are given by the elements of $\mathbf{W}(\mathbf{\Omega})$. These couplings include a normalization that causes the effect of the damage represented by $\mathbf{W}(\mathbf{\Omega}^\star)$, with elements given by Eq. (\ref{w_ij_Omega_explicit}), to incorporate two combined effects: the reduction of the weight on a specific link and the redistribution of the lost function by other connections in the network. In order to explore the existence of antifragility in coupled oscillators emerging solely from the reduction of the coupling weight $\mathbf{\Omega}$, this section analyzes synchronization processes defined by a Kuramoto model with non-normalized coupling weights.
\\[2mm]
We start with the Kuramoto model for identical oscillators with non-normalized couplings  \cite{Eraso-Hernandez_2023}
\begin{equation}\label{kuramoto_Omegaij}
	\frac{d\theta_i(t)}{dt}=\sum_{j=1}^{N
	}\Omega_{ij}\sin[\theta_j(t)-\theta_i(t)],
\end{equation}
choosing the weights $\mathbf{\Omega}=\mathbf{A}$ in a system that reaches synchronization up to a given threshold value $r$, for which the time $\tau_0$ is defined.
\\[2mm]
Similarly, synchronization is studied in a modified system with weights $\mathbf{\Omega}^\star$, where damage is introduced in the link $a \to b$: $\Omega^\star_{ab}=(1-\beta) A_{ab}$ (the rest of the entries coincide with the original weights in $\mathbf{\Omega}$). This new system evolves according to the following set of differential equations
\begin{equation}\label{kuramoto_Omegastarij}
	\frac{d\theta_i(t)}{dt}=\sum_{j=1}^{N
	}\Omega^\star_{ij}\sin[\theta_j(t)-\theta_i(t)].
\end{equation}
In this case, $\tau(\beta)$ denotes the time required to reach the order parameter $r$. From the information given by $\tau_0$ and $\tau(\beta)$, obtained under the same random initial condition for the phases (chosen from a uniform distribution), we compute the functionality $\mathcal{F}_\beta = \frac{\tau_0}{\tau(\beta)}$ and the coefficient $\Lambda^{(\mathrm{KM-NNC})}$, which measures the response to damage for infinitesimal variations in the link $a \to b$ and is calculated numerically as
\begin{figure*}[t!]
	\centering
	\includegraphics*[width=0.95\textwidth]{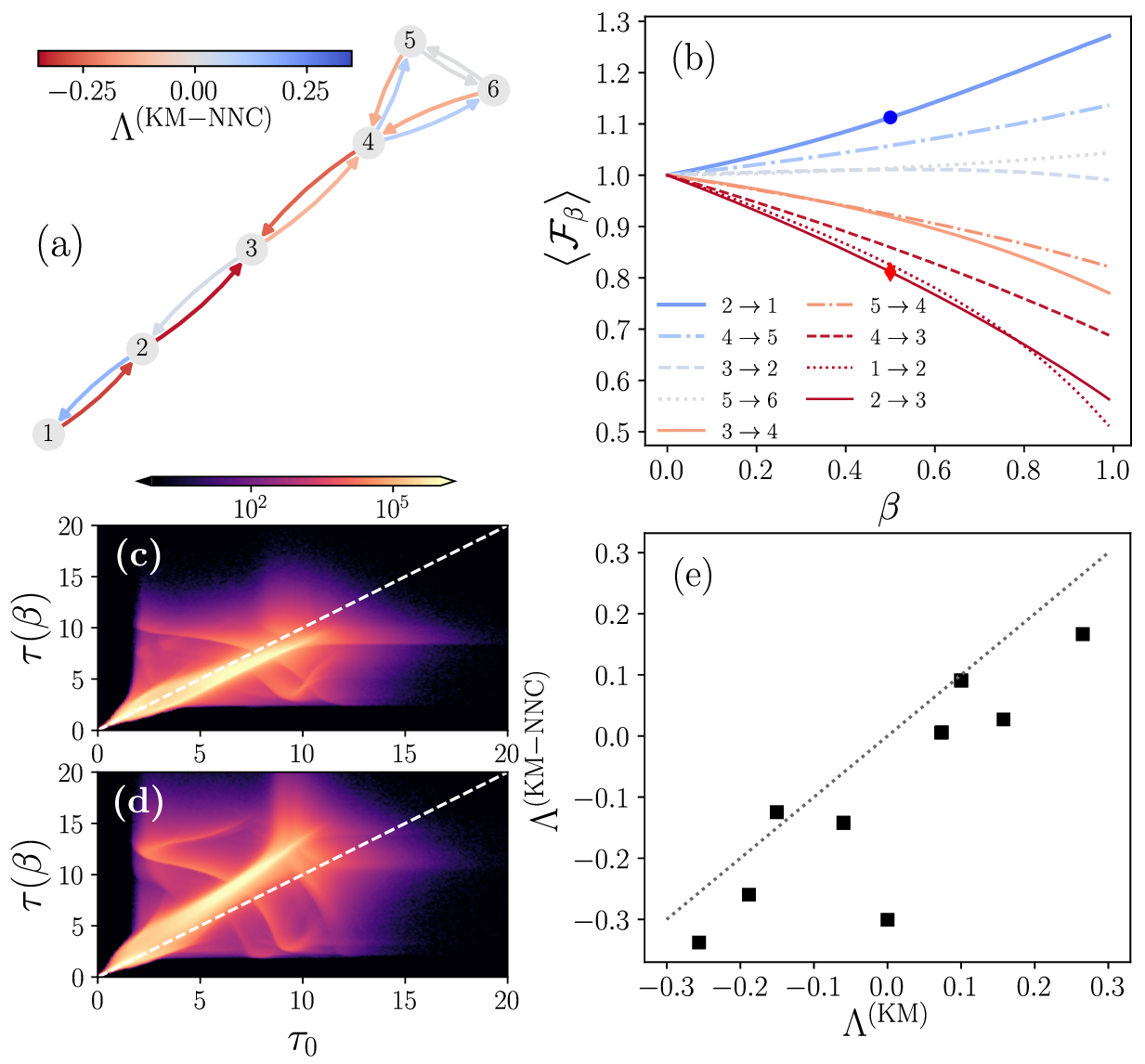}
	\vspace{-5mm}
	\caption{Antifragility in the response to damage in the Kuramoto model with non-normalized couplings in a lollipop graph. (a) Representation of the analyzed graph, where the edges are colored according to the values of $\Lambda^{(\mathrm{KM-NNC})}$. (b) Ensemble average of the functionality $\langle \mathcal{F}_\beta \rangle$ as a function of $\beta$ for all edges in the graph shown in (a). (c)-(d) Two-dimensional histograms of the pairs $(\tau_0, \tau(\beta))$ for $\beta=0.5$, calculated for the modified edges: (c) $2 \to 1$, and (d) $2 \to 3$  (each dashed line depicts the relation $\tau(\beta)=\tau_0$). (e) Comparison between $\Lambda^{(\mathrm{KM})}$ from Eq. (\ref{Lambda_edge_numerical}) for the Kuramoto model in Eq. (\ref{kuramoto_wij}) and $\Lambda^{(\mathrm{KM-NNC})}$ from Eq. (\ref{Lambda_edge_KM-NNC}) for the model with non-normalized couplings. The analysis includes all edges of the graph shown in (a). See the main text for further details.}
	\vspace{2mm}
	\label{Fig_7}
\end{figure*}
\begin{equation}\label{Lambda_edge_KM-NNC}
	\Lambda^{(\mathrm{KM-NNC})}\approx \frac{1}{\Delta \beta}\left[\langle \mathcal{F}_{\Delta \beta}\rangle-1\right]
\end{equation}
for a small value $\Delta \beta$.
\\[2mm]
In order to illustrate the effect of damage in synchronization processes where the couplings are not normalized, in Fig.~\ref{Fig_7} we show different results for the study of the lollipop graph previously analyzed in Fig.~\ref{Fig_1}. All the numerical values are obtained using the synchronization threshold $r=0.99$. 
\\[2mm]
Figure~\ref{Fig_7}(a) depicts the analyzed graph. The value $\Lambda^{(\mathrm{KM-NNC})}$ is obtained for each edge using Eq.~(\ref{Lambda_edge_KM-NNC}) with $\Delta \beta=0.01$ and $10^7$ realizations to compute the ensemble averages (the results are encoded in the colorbar). Figure~\ref{Fig_7}(b) shows the ensemble average $\langle \mathcal{F}_\beta \rangle$ as a function of $\beta \in [0,1)$ for all directed edges in the graph (the cases $4 \to 6$, $6 \to 4$, and $6 \to 5$ are not included due to their symmetry with other edges). We use $10^6$ realizations to compute each value of $\langle \mathcal{F}_\beta \rangle$. The curves were ordered according to the corresponding values of $\Lambda^{(\mathrm{KM-NNC})}$, from the largest to the smallest. On the other hand, Figs. \ref{Fig_7}(c)-(d) show the statistical analysis of the pairs $(\tau_0, \tau(\beta))$, with $\beta=0.5$ and $10^9$ realizations, for the edges $2\to 1$ [in Fig. \ref{Fig_7}(c)], which exhibits the largest antifragile response, and $2\to 3$ [in Fig. \ref{Fig_7}(d)], which corresponds to the most fragile response. The respective functionality of these cases, for $\beta=0.5$, is indicated with a marker in Fig. \ref{Fig_7}(b).
\\[2mm]
For the antifragile response shown in Fig. \ref{Fig_7}(c), a higher density of points is observed below the dashed line, corresponding to scenarios where $\tau_0 > \tau(\beta)$. In contrast, in Fig. \ref{Fig_7}(d), the response of the system is fragile, with a more significant number of random initial conditions producing  $\tau_0 < \tau(\beta)$. Additionally, it is worth highlighting that the patterns observed in Figs. \ref{Fig_7}(c)-(d) are similar to those found in Figs. \ref{Fig_1}(c)-(f), showing regions with a higher density of time pairs $(\tau_0, \tau(\beta))$ associated with specific responses of the system under different initial conditions. 
\\[2mm]
Figure~\ref{Fig_7}(e) compares the results obtained for $\Lambda^{(\mathrm{KM-NNC})}$ from Eq.~(\ref{Lambda_edge_KM-NNC}) with those obtained for $\Lambda^{(\mathrm{KM})}$ from Eq.~(\ref{Lambda_edge_numerical}), using the values corresponding to all the links of the graph analyzed in Fig.~\ref{Fig_7}(a). The results reveal that, when the dynamics are governed by non-normalized couplings, a new classification of the links emerges based on their response to damage. Moreover, it is important to highlight that, in most cases, the inequality $\Lambda^{(\mathrm{KM-NNC})} < \Lambda^{(\mathrm{KM})}$ holds, as shown by the points located below the dashed line representing $\Lambda^{(\mathrm{KM-NNC})} = \Lambda^{(\mathrm{KM})}$. The only exception to this behavior corresponds to the link $3 \to 4$, where the application of infinitesimal damage results in $\Lambda^{(\mathrm{KM-NNC})} > \Lambda^{(\mathrm{KM})}$. 
\\[2mm]
The results presented in Fig.~\ref{Fig_7} provide evidence of a case where the effect of antifragility emerges when the coupling weights are not normalized. This observation suggests that antifragility is an inherent property of synchronization processes, which can be further enhanced in systems where, after damage to a specific link, other parts of the system compensate by absorbing the lost functionality. Finally, the findings presented in this section demonstrate the general applicability of the proposed methods for characterizing antifragility and fragility in systems of coupled oscillators that reach synchronization, i.e., in cases where it is possible to compute the times $\tau_0$ and $\tau(\beta)$, which remain finite for a given synchronization threshold $r$.
\section{Conclusions}
\label{Sec_Conclusions}
In this research, we introduce a mathematical framework to evaluate the impact of damage, understood as the reduction of weight in a specific link, in oscillator systems coupled through weighted networks. We analyze the functionality $\mathcal{F}_\beta$, which quantifies the effect of weight modifications on a link when the system's global function is to achieve in a finite time global synchronization of coupled oscillators starting from random initial phases. The ensemble average $\langle \mathcal{F}_\beta \rangle$ provides a combined description of both the network structure and the synchronization process, while its derivative evaluated at $\beta \to 0$ allows to calculate $\Lambda^{(\mathrm{KM})}$ that measures the effect of infinitesimal weight variations in a specific link. If $\Lambda^{(\mathrm{KM})} > 0$, the system exhibits antifragility, meaning that the reduction in the weight of a link enhances its ability to achieve synchronization. Conversely, $\Lambda^{(\mathrm{KM})} < 0$ indicates fragility, as the infinitesimal damage leads to a decrease in synchronization capacity. The case $\Lambda^{(\mathrm{KM})} = 0$ corresponds to situations where small weight modifications in a specific link have no effect on synchronization times.
\\[2mm]
By implementing this methodology through the numerical solution of the Kuramoto model for identical oscillators and leveraging the computational power of graphics processing units (GPUs) to parallelize the exploration of multiple initial conditions, we efficiently evaluated $\langle \mathcal{F}_\beta \rangle$ and $\Lambda^{(\mathrm{KM})}$ for different types of graphs. In particular, we analyzed the impact of damage on all edges of 109 non-isomorphic graphs with $N=6$ nodes. In addition to $\langle \mathcal{F}_\beta \rangle$ and $\Lambda^{(\mathrm{KM})}$, we analyzed other measures, including $\Lambda^{(\mathrm{linear})}$ for the linear approximation of the Kuramoto model and $\Lambda^{(\xi_2)}$, which uses only the real part of the second smallest eigenvalues $\xi_2(\mathbf{A})$ and $\xi_2(\mathbf{\Omega}^\star)$ for infinitesimal weight variations. These quantities were compared with $\Lambda^{(\mathrm{RW})}$, introduced in Ref. \cite{Eraso_PRE_2024} to study antifragility in systems whose global function is the efficient transport between nodes using random walks on weighted networks. Our results for graphs with $N=6$ nodes indicate that the classification of fragility/antifragility based on $\Lambda^{(\mathrm{KM})}$ does not directly correlate with that obtained using $\Lambda^{(\mathrm{RW})}$. This finding highlights that the impact of damage depends not only on the network structure and the affected link but also on the specific dynamical process governing the global function of the system.
\\[2mm]
The proposed framework is general and can be applied to oscillator systems that achieve synchronization within a finite time considering different coupling matrices, heterogeneous oscillators, or extensions of the Kuramoto model. The methods introduced provide a foundation for a broader understanding of the emergence of antifragility and the impact of damage on dynamical processes in complex systems.

\section{Appendix: Antifragility of transport on networks with damage }
\label{Appendix_RW}
\label{Appendix_RW}
In this appendix, we summarize the main concepts and definitions used to evaluate the effect of infinitesimal variations in random walk dynamics (see Ref. \cite{Eraso_PRE_2024} for a detailed discussion). The process is modeled as a Markovian random walker on a connected, undirected network with $N$ nodes labeled $i = 1, \ldots, N$. The walker begins at $t = 0$ from node $i$ and, at each step, hops to a new node according to the transition probabilities defined by the matrix $\mathbf{W}(\mathbf{\Omega})$, whose elements are given by $w_{i \to j}(\mathbf{\Omega}) \equiv \Omega_{ij} / \mathcal{S}_i$, where $\mathcal{S}_i = \sum_{l=1}^N \Omega_{il}$ is the generalized strength of node $i$ \cite{ReviewJCN_2021}. The standard random walk strategy is defined using $\Omega_{ij} = A_{ij}$, meaning that the walker moves from a node to one of its nearest neighbors with equal probability \cite{MasudaPhysRep2017}. In this case, we denote the transition matrix as $\mathbf{W}$, specifically $\mathbf{W}(\mathbf{A})$.
\\[2mm]
The capacity of this process to explore the entire structure is quantified by a global time, $\tau^{(\mathrm{RW})}_0$, defined in terms of the eigenvectors and eigenvalues of $\mathbf{W}$. Let the eigenvalues of $\mathbf{W}$ be denoted by $\lambda_l$ (with $\lambda_1 = 1$) and its right and left eigenvectors by $\left|\phi_l\right\rangle$ and $\left\langle\bar{\phi}_l\right|$, respectively, for $l = 1, 2, \ldots, N$. These eigenvectors satisfy the orthogonality condition $\left\langle\bar{\phi}_l|\phi_m\right\rangle = \delta_{lm}$ and the completeness relation $\sum_{l=1}^N \left|\phi_l\right\rangle \left\langle\bar{\phi}_l\right| = \mathbf{I}$, where $\mathbf{I}$ is the $N \times N$ identity matrix \cite{ReviewJCN_2021}. These quantities can be determined either numerically or analytically, depending on the case. Using this information, the global time is defined as $\tau^{(\mathrm{RW})}_0 \equiv \frac{1}{N} \sum_{j=1}^N \tau_j(0)$, where \cite{ReviewJCN_2021}
\begin{equation}\label{tau_j0}
	\tau_j(0)=\frac{1}{\left\langle j|\phi_1\right\rangle \left\langle\bar{\phi}_1|j\right\rangle}\sum_{\ell=2}^N \frac{1}{1-\lambda_\ell}\left\langle j|\phi_\ell\right\rangle \left\langle\bar{\phi}_\ell|j\right\rangle.
\end{equation}
Let us now define a modified dynamics with a reorganization of the transition probabilities in $\mathbf{W}$ due to a reduction of the capacity of transport (damage) in the link $a \to b$ connecting $a$ to $b$ in the network. The transition matrix $\mathbf{W}(\mathbf{\Omega}^\star)$ defines an ergodic process for  $0\leq \beta<1$. In particular, due to the normalization, $\mathbf{W}(\mathbf{\Omega}^\star)$ and the standard random walk defined by $\mathbf{W}$ differ in the $a$-th row. Also, it is possible, at least numerically, to obtain a global time denoted as $\tau^{(\mathrm{RW})}(\beta)$ similar to $\tau^{(\mathrm{RW})}_0$, but now using the eigenvalues and eigenvectors of  $\mathbf{W}(\mathbf{\Omega}^\star)$. In terms of these quantities, it is defined the global functionality for the random walk dynamics \cite{Eraso_PRE_2024}
\begin{equation}\label{F_beta_edge_RW}
	\mathcal{F}^{(\mathrm{RW})}_\beta\equiv\frac{\tau^{(\mathrm{RW})}_0}{\tau^{(\mathrm{RW})}(\beta)},
\end{equation}
that measures the global response of the system  to the modification in the edge $a \to b$. In particular, if we consider the effect of an infinitesimal reduction in the capacity of the link, it is convenient to define \cite{Eraso_PRE_2024}
\begin{equation}\label{Lambda_edge_RW}	\Lambda^{(\mathrm{RW})}\equiv\frac{d\mathcal{F}^{(\mathrm{RW})}_\beta}{d\beta}\Big|_{\beta\to 0}
\end{equation}
Thus, $\Lambda^{(\mathrm{RW})} > 0$ indicates antifragility, while $\Lambda^{(\mathrm{RW})} < 0$ reflects a reduction in the transport capacity due to infinitesimal damage to the link. A value of $\Lambda^{(\mathrm{RW})} = 0$ occurs when an infinitesimal variation in the weight of a specific link does not alter $\tau^{(\mathrm{RW})}(\beta)$.
\section*{References}

\providecommand{\noopsort}[1]{}\providecommand{\singleletter}[1]{#1}%
\providecommand{\newblock}{}

\end{document}